\documentclass[twocolumn,
superscriptaddress,
prb,
showpacs,
preprintnumbers,
amsmath,amssymb]{revtex4}

\usepackage{graphicx,epsf,epsfig,ulem}
\usepackage{epsfig}
\usepackage{epstopdf} 
\usepackage{bm}
\usepackage{subfigure}
\usepackage{color}
\usepackage{amsmath}
\begin{document}

\title{Engineering inter-qubit exchange coupling between donor bound electrons in silicon}
\textbf{}

\author{Yu E. Wang*}
\affiliation{Network for Computational Nanotechnology, Purdue University, West Lafayette, IN 47907, USA}

\author{Archana Tankasala}
\affiliation{Network for Computational Nanotechnology, Purdue University, West Lafayette, IN 47907, USA}

\author{Lloyd C. L. Hollenberg}
\affiliation{Centre for Quantum Computation and Communication Technology, School of Physics, University of Melbourne, VIC 3010, Australia}

\author{Gerhard Klimeck}
\affiliation{Network for Computational Nanotechnology, Purdue University, West Lafayette, IN 47907, USA}

\author{Michelle Y. Simmons}
\affiliation{Centre for Quantum Computation and Communication Technology, School of Physics, University of New South Wales, Sydney, NSW 2052, Australia}

\author{Rajib Rahman*}
\affiliation{Network for Computational Nanotechnology, Purdue University, West Lafayette, IN 47907, USA}

\date{\today}

\begin{abstract}
We investigate the electrical control of the exchange coupling (J) between donor bound electrons in silicon with a detuning gate bias, crucial for the implementation of the two-qubit gate in a silicon quantum computer. We find the asymmetric 2P-1P system provides a highly tunable exchange-curve with mitigated J-oscillation, in which 5 orders of magnitude change in the exchange energy can be achieved using a modest range of electric field for 15 nm qubit separation. Compared to the barrier gate control of exchange in the Kane qubit, the detuning gate design reduces the demanding constraints of precise donor separation, gate width, density and location, as a range of J spanning over a few orders of magnitude can be engineered for various donor separations. We have combined a large-scale full band atomistic tight-binding method with a full configuration interaction technique to capture the full two-electron spectrum of gated donors, providing state-of-the-art calculations of exchange energy in 1P-1P and 2P-1P qubits.    

\end{abstract}


\maketitle 
\section{Introduction}
Quantum computation has the potential to solve classes of problems that are intractable to classical computing, such as large integer factorization which is important in cryptography \cite{Ekert}. A quantum computer based on silicon can take advantage of the existing device fabrication platform of the semiconductor industry and the exceptional spin coherence times measured to be in excess of 35 s in recent experiments \cite{Muhonen}. 
Donor qubits in silicon are promising candidates for a spin based quantum computer as they have exceptionally long $T_1$ \cite{Morello, Buch, Hsueh} and $T_2$ times \cite{Pla, Muhonen, Saeedi} and offer both electron and nuclear spins for encoding quantum information \cite{Pla, Kane, Pla1, Saeedi} utilizing the already existing and commonly used silicon device technology. 

While single qubits in silicon have been demonstrated with both electronic and nuclear spins of donors \cite{Pla1, Pla}, the next biggest challenge is to demonstrate two-qubit gates based on exchange interaction. Ideally, the exchange coupling J in a two-qubit gate needs to be tuned electrically by several orders of magnitude between an 'Off' and an 'On' state within a small and realizable bias range. To achieve this, the popular Kane architecture uses a J-gate between two phosphorus donors to tune the J-coupling and A-gates to control individual qubits (Fig. 1(a)) \cite{Kane}. However, such an architecture leads to a high gate density, ultra small gate widths, electrical cross-talk between gates, and precise donor positioning relative to gates. Moreover, the tunability of the exchange coupling is limited both by the field range the J-gate can produce and the field ionization of the electrons to the surface. Calculations have also shown that the J-coupling oscillates as a function of donor separation due to crystal momentum states \cite{Koiller}, and is therefore sensitive to atomic scale placement errors. All these issues cause severe constraints in the implementation of a two-qubit gate in donors.




\begin{figure}[htbp]
\center\epsfxsize=3.4in\epsfbox{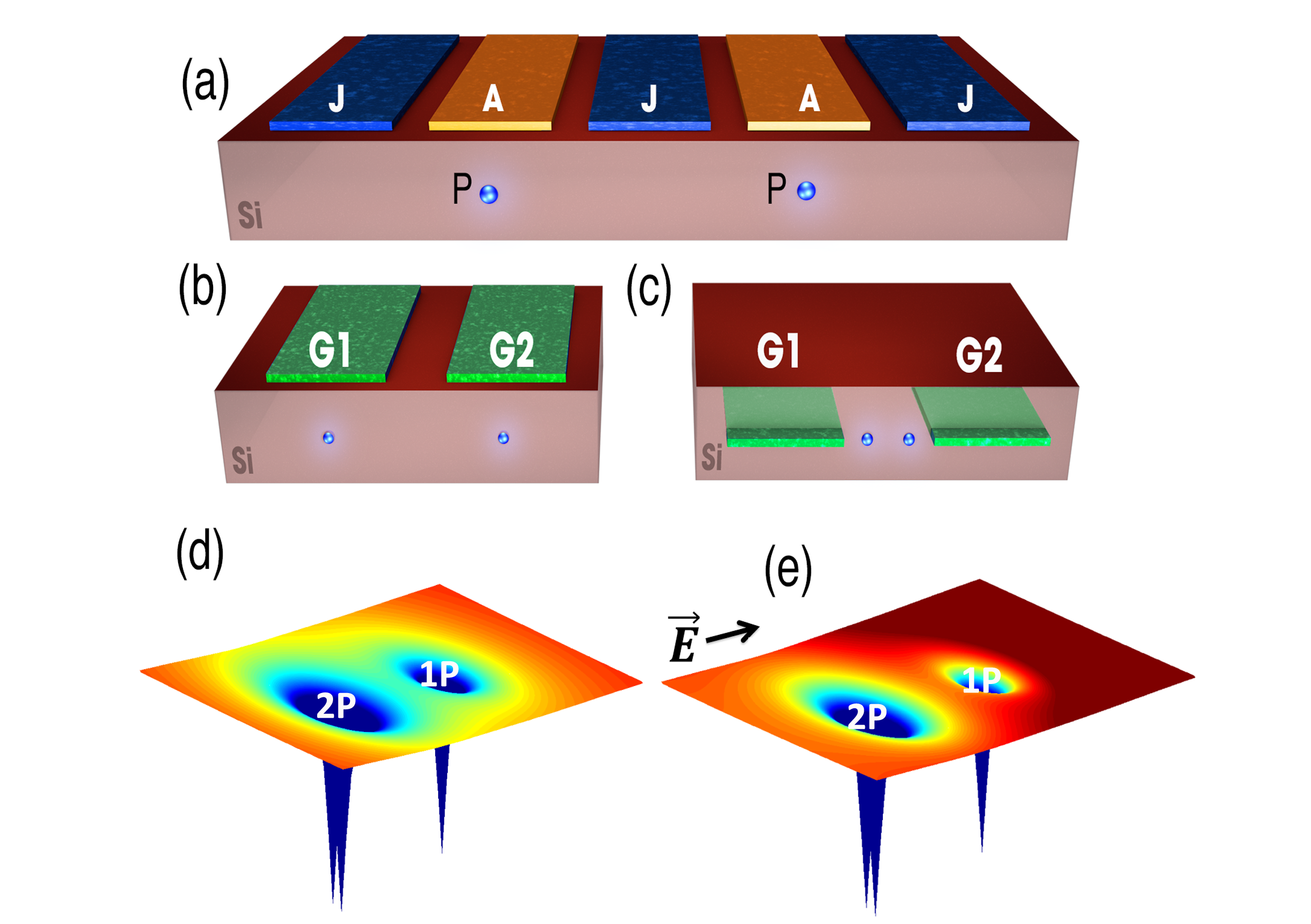}
\caption{ Control of exchange in donor based qubits separated by $\sim$15 nm. (a) Kane's two-qubit donor device with J-gate \cite{Kane}. (b) A MOS device with top detuning gates. (c) An STM patterned device with in-plane detuning gates. (d),(e) schematic potential energy profiles of the two-qubit system without and with the detuning electric field $\vec{E}$, showing a tilt in potential energy along the separation direction in (e).}
\vspace{0cm}
\label{fig_devices}
\end{figure}

In this work, we introduce an alternate design for an exchange gate in a two-qubit donor system, which allows flexibility in device fabrication and in tuning the exchange coupling. In principle, this new design can 1) eliminate the need for additional J-gates between the donors, 2) function with a range of donor separations, 3) provide a 5 order of magnitude J-tunability within a modest E-field range of $\sim$3 MV/m and lowered 'Off' state exchange, and 4) mitigate the J-oscillations with donor separations. This design can also benefit from improved addressability \cite{Buch} and longer $T_1$ times \cite{Hsueh}, and can be used in conjunction with the two-qubit scheme proposed in Ref \cite{Rachpon}. This design can therefore overcome some of the experimental obstacles for realizing a two-qubit gate in silicon. To perform these state-of-the-art calculations of exchange energy in donor qubits, we have combined atomistic full band electronic structure calculations, electrostatic device simulations, and a two-electron full configuration interaction (FCI) technique over a lattice of 1.7 million atoms. The calculations therefore describe the detailed two-electron spectrum of donor qubits over large E-field ranges accounting for both crystal effects and electron-electron exchange and correlation effects. 

By studying double quantum dots (DQD) based on donor clusters in silicon \cite{Weber_DQD}, we find that an asymmetric 2P-1P system outperforms the symmetric 1P-1P system in exchange controllability with detuning gates. Analogous to exchange-tuning in a DQD \cite{Petta, Rahman_DQDCI}, we envision a (1,1) to (2,0) charge transition \cite{charge_config_note} in both 1P-1P and 2P-1P qubits as a function of a lateral electric field which provides the energy detuning. As the electron bound to one donor is pulled to the other by the electric field, the exchange coupling can be engineered from a small value in the (1,1) state to a large value in the (2,0) state due to the large spatial overlap of the wavefunctions in the latter. Such a detuning field can be applied from either top gates (see Fig. 1 (b)) in a metal-oxide-semiconductor (MOS) device or from in-plane gates realized by scanning tunneling microscope (STM) based lithography \cite{Weber_DQD}. Placed on either side of the donor qubits, the detuning gates eliminate the need for a sensitive tunnel barrier control by the J-gate. Instead, this design realizes a tilt in the potential landscape of the two qubits, as shown in Fig. 1(d) and (e), and therefore relaxes the more stringent engineering requirements of donor separations and gate widths of the Kane architecture, leading to a reduced gate density of the computer. 


\section{Methods}
Previous works have calculated the exchange energy between two donor-bound electrons in silicon as a function of separation $R$ using the effective mass approximation (EMA) and the Heitler-London (HL) formalism \cite{Koiller,Wellard}, which is valid in the regime of small wavefunction overlap. Control of exchange with the Kane J-gate was also calculated in \cite{Wellard,Wellard_EM} from EMA based HL method. However, this method becomes inaccurate at modest gate biases when the wavefunction overlap increases. The HL method also ignores contributions of doubly excited configurations that increase with reduced separations or increasing fields. More recently, the exchange coupling was calculated in a 1P-1P system from EMA based molecular orbital and configuration interaction approaches for separation distances larger than 7.5 nm \cite{Kettle} and modest electric fields \cite{Gonzalez-Zalba}. All these works are based on the Kohn-Luttinger form of the donor wavefunctions \cite{Kohn}, which provides a very specific solution to the two-electron problem, and cannot provide a full description of the (1,1) to (2,0) charge transition in which strong Stark effect causes mixing of the lowest states with many excited states \cite{Rahman_orbital_stark_effect}. The atomistic configuration interaction method used here goes beyond these approximations.


In this work, we employ a large scale atomistic tight-binding (TB) method to simulate the electronic structure of P donors in silicon. The full Hamiltonian of silicon and donor atoms is represented by the $sp^{3}d^{5}s^{*}$ atomistic basis with nearest neighbor interactions \cite{Klimeck_ted}. This method expresses the single electron wavefunction as a linear combination of the atomic orbitals. For a silicon unit cell, the model reproduces the full band structure including the correct conduction band and valence band extrema, effective masses, and band gap \cite{Boykin}. A donor is modeled by the Coulomb potential of a positive charge screened by the silicon dielectric constant, and an onsite cut-off potential $U_0$ adjusted to reproduce the ground state binding energy of 45.6 meV below the conduction band for a P donor in Si \cite{Rahman_prl}. It has been shown in earlier works that this model captures the central-cell correction, and produces correct wavefunction symmetries and energy splittings of the donor states \cite{Ahmed}. It also captures the valley-orbit interaction \cite{Rahman_VO}, the Stark shift of the donor orbitals \cite{Rahman_orbital_stark_effect}, and the computed ground state wavefunction also showed excellent agreement with real and momentum space images of the donor obtained by STM experiments \cite{Salfi}. Using this model, the single electron molecular orbitals of the two donors are obtained by,

\begin{equation}
  H_{1,2}=H_0+H_L+H_R+ e\vec{E}\cdot\vec{r}, \label{eq1}
\end{equation}


\noindent where $H_{1,2}$ represents the single electron Hamiltonian for electrons 1 or 2, $H_0$ is the crystal Hamiltonian of millions of Si atoms, $H_L$ and $H_R$ are the left and right donor potential energy respectively, and the last term is the potential energy of a detuning electric field $\vec{E}$. The eigen problem $H_i \Psi= \epsilon\Psi$ is solved by a parallel block Lanczos algorithm to obtain a set of molecular orbitals of the P-P$^+$ system, $\lbrace \Psi_1, \Psi_2, ...., \Psi_m \rbrace$ corresponding to energies $\lbrace \epsilon_1, \epsilon_2, ...., \epsilon_m \rbrace$ where the index goes from the ground state 1 to the excited state $m$. 



Using the $m$ molecular orbitals, all possible two-electron Slater determinants (SD) are constructed to represent non spin-adapted anti-symmetric electron configurations of electrons occupying molecular orbitals $i$ and $j$ as $SD_{ij}=\frac{1}{\sqrt{2}}[\Psi_i(\vec{r}_1) \Psi_j(\vec{r}_2)-\Psi_j(\vec{r}_1)\Psi_i(\vec{r}_2)]$, where $\vec{r}_1$ and $\vec{r}_2$ are the coordinates of electron 1 and 2. We employ a full configuration interaction (FCI) scheme, which means that we include all possible single and double excitations of the system for the two-electron problem. We found that about 12 single electron molecular orbitals (24 spin states) are needed to achieve convergence, which leads to the number of SDs being $C^{2}_{24}$. The two-electron Hamiltoinian is given as,
 
\begin{equation}
  H_{2e}=H_1+H_2+\frac{e^2}{4\pi\epsilon(|\vec{r}_1-\vec{r}_2|)}\frac{1}{|\vec{r}_1-\vec{r}_2|}, \label{eq2}
\end{equation}

\noindent where the last term denotes the Coulomb interaction between electrons 1 and 2 with the dielectric function being $\epsilon(|\vec{r}_1-\vec{r}_2|)$. If this term is evaluated in the basis of the SDs, the familiar Coulomb (C) and exchange (K) integrals are obtained, as follows,

\begin{equation}
 \langle SD_{ij}|\frac{e^2}{4\pi\epsilon(|\vec{r}_1-\vec{r}_2|)}\frac{1}{|\vec{r}_1-\vec{r}_2|}|SD_{pq}\rangle=C_{ij,pq}-K_{ij,pq},\label{eq3}
\end{equation}
where $i, j, p, q \le m$. The first and the second terms on the right hand side are the $C_{ij,pq}$ and $K_{ij,pq}$ integrals:
\begin{equation}
C_{ij,pq}=\iint d\vec{r}_1d\vec{r}_2\Psi^{*}_{i}(\vec{r}_1)\Psi^{*}_{j}(\vec{r}_2)\frac{1}{|\vec{r}_1-\vec{r}_2|}\Psi_{p}(\vec{r}_1)\Psi_{q}(\vec{r}_2), \label{eq4}
\end{equation}
\begin{equation}
K_{ij,pq}=\iint d\vec{r}_1d\vec{r}_2\Psi^{*}_{i}(\vec{r}_1)\Psi^{*}_{j}(\vec{r}_2)\frac{1}{|\vec{r}_1-\vec{r}_2|}\Psi_{q}(\vec{r}_1)\Psi_{p}(\vec{r}_2). \label{eq5}
\end{equation}
The solution of Eq. \ref{eq2} in the basis of SDs yields the total energies and wavefunctions of the two-electron system, which comprises of spin singlets and triplets. The wavefunctions appear as linear combinations of SDs with coefficients $c_{ij}$, given as $\Psi_{2e}=\sum_{i,j}c_{ij}SD_{ij}$, where $i, j \le m$ and $i\ne j$. 

 
The FCI technique is an exact way to solve the multi-electron problem only limited by the number of one electron basis functions used. The method diagonalizes the multi-electron Hamiltonian in the basis of all Slater determinants (SD) constructed from the single electron states of the system. Each SD represents a multi-electron anti-symmetric wavefunction for a particular arrangement of the electrons among the basis orbitals. In addition to the ground state, the method also captures the excited multi-electron spectra, including exchange and correlation energies \cite{Sherrill}. This method has been successful in solving the challenging problem of the $D^-$ state (the two electron state of a single donor) without any fitting parameters and providing a charging energy of 45 meV \cite{Archana} compared to the 44 meV experimental value \cite{Rahman_CE}. 
The exchange energy in the two-electron case can be obtained from the difference between the lowest triplet and singlet energies, $E_T$ and $E_S$ respectively, as $\Delta E_{ST}=E_T-E_S$. 

\begin{figure}[htbp]
\center\epsfxsize=3.4in\epsfbox{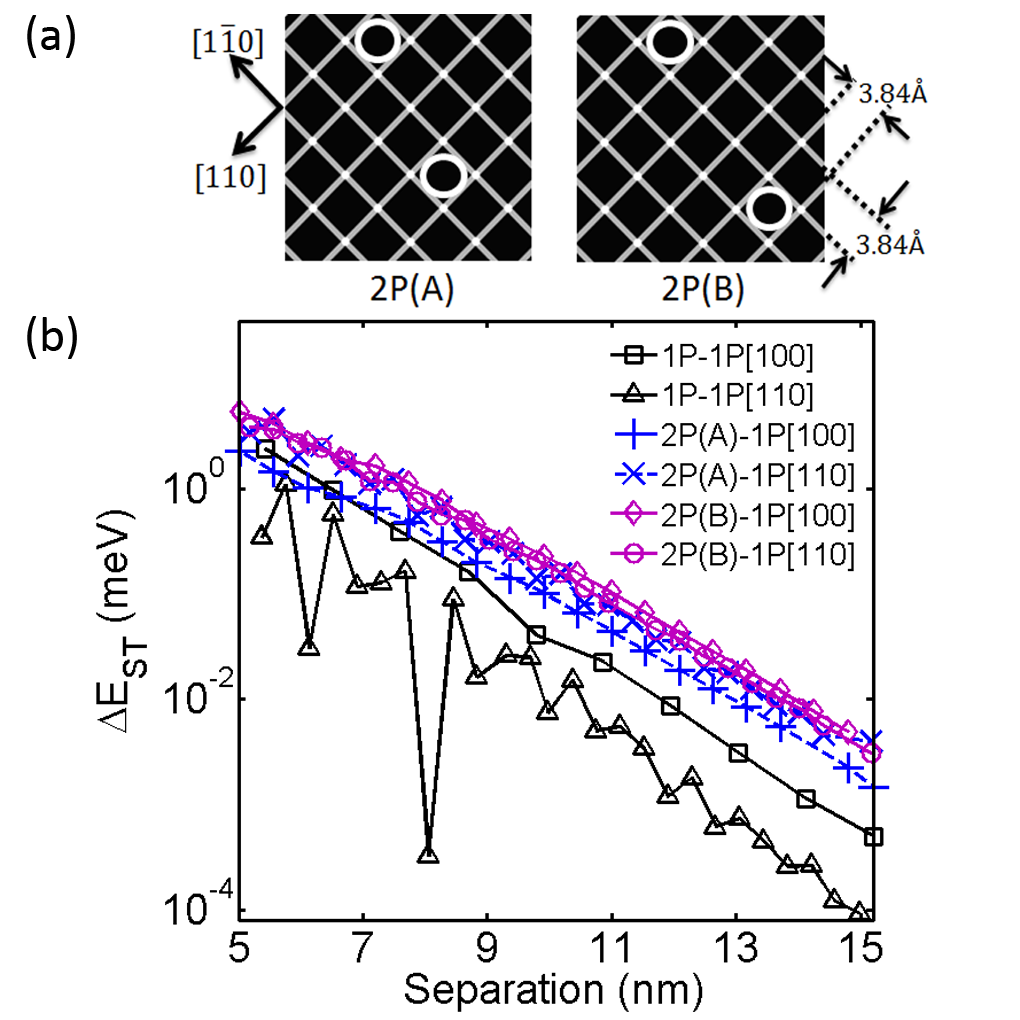}
\caption{Separation dependence of exchange energy. (a) Two studied 2P cases with different donor locations on a (001) plane in silicon. (b) The exchange energy of 1P-1P and 2P-1P qubits in silicon as a function of inter-qubit separation along the [100] and [110] crystallographic axes without any applied electric fields.}
\vspace{0cm}
\label{fij_r}
\end{figure}
\section{Results and discussions}
\subsection{A. Exchange Energy vs inter-qubit separation}
In Fig. 2(b) we show the computed exchange energy between two donor bound electrons as a function of donor separation along two high symmetry crystallographic directions [100] and [110]. We consider two cases: 1) a symmetric two-qubit system of 1P-1P \cite{Kane} (black triangles and squares), and 2) an asymmetric two-qubit system of 2P-1P along the lines of Refs \cite{Buch, Weber_DQD}, in which we consider two representative locations of the donors inside the 2P cluster, shown in Fig. 2(a) as 2P(A) and 2P(B). The exchange energy as function of 2P-1P separation for these two spatial configurations of the 2P cluster are labeled as 2P(A) (blue curves) and 2P(B) (magenta curves) in Fig. 2(b). In all the cases, the exchange energy decreases exponentially with increasing separation in both the [100] and the [110] directions, attaining a few meV for separation $\sim$5 nm and a few $\mu$eV or less for $\sim$15 nm. This is expected since the wavefunction overlap between the qubits decreases as the separation increases. 

The exchange energy of the 2P-1P qubits are found to be slightly larger than the 1P-1P qubits even without an electric field. This is somewhat counter-intuitive considering the fact that a 2P system provides a stronger confinement for the electron and should minimize the overlap with the 1P electron. However, due to the asymmetry of the confinement potential, the 2P cluster is already slightly detuned towards the (2,0) state with larger wavefunction overlap, resulting in a larger exchange. The exchange can be reduced below the 1P-1P value in this case by applying a negative detuning bias, as shown later.

The exchange energy of the 1P-1P qubits in the [110] direction (black triangles) also exhibits oscillations with the donor separation. This has been observed in earlier works based on the effective mass HL method \cite{Koiller,Wellard}, and has been attributed to the interference between the Bloch components of the wavefunctions of the six conduction band valleys of silicon. The location of the atom sites in the zincblende lattice dictates that the donor separation in [100] can be increased by at least an unit cell of lattice constant $a_0$. However, in [110] the minimum unit of separation increases by $a_0/ \sqrt{2}$, which is a separation increment of $a_0/2$ in each [100] and [010] directions.  

It is interesting to note that the J-oscillations are strongly mitigated in all the 2P-1P cases along both [100] and [110] separation. In fact, we considered the two 2P cluster configurations, 2P(A) and 2P(B), to verify that this is not a one-off case. In 2P(A), the intra-cluster P donors are separated by $a_0$ times an integer in both [100] and [010], and in 2P(B) by $\frac{a_0}{2}$ times an odd integer. These two cases are thus representatives of the donors separated by a whole unit cell and a half unit cell. Relative to this 2P cluster, if the 1P distance is varied along [100] or [110], the J-oscillations should appear in one case in analogy with the 1P-1P system. However, as shown in Fig 2(b), we observe strongly suppressed J-oscillations in all the 2P-1P cases possibly due to a randomized interference pattern between the three donor wavefunctions. 
Hence, the 2P-1P unit is a more fault-tolerant system for fabrication and control than the 1P-1P unit.

\subsection{B. Exchange energy vs detuning electric-field}
Now that we have explored the range of exchange energies that can be accessed in donor qubits for various donor separations, we investigate the detuning E-field control of exchange in the 1P-1P and 2P-1P systems. Fig. 3 shows the exchange energies as a function of the detuning field for 10 nm and 15 nm qubit separations respectively. We consider an electric field range of -2 to 2 MV/m, consistent with typical E-fields realized in STM patterned donor devices. In general, the exchange curves show a transition from a low value to a large value for a charge transition from (1,1) to (2,0) for the 2P-1P curves. This is because when the two electrons are localized on separate qubits, the spatial overlap of the wavefunction is small, and so is the J coupling. Since a 2P-1P system is slightly detuned toward the 2P cluster at zero detuning, a negative detuning field further decreases the wavefunction overlap so that the exchange energy is decreased. A positive field pulls both the electrons gradually to one qubit, such that the spatial overlap increases, and so does the exchange. In the extreme (2,0) regime when both electrons are on the left qubit, the detuning field does not affect the overlap anymore, and the exchange energy levels off. All these regimes are clearly visible in the 2P-1P exchange curve for 15 nm separation, in which the exchange splitting varies from 48 neV to 5.5 meV over a field range of -2 to 0.9 MV/m, resulting in over five orders of magnitude tunability. 

\begin{figure}[htbp]
\center\epsfxsize=3.4in\epsfbox{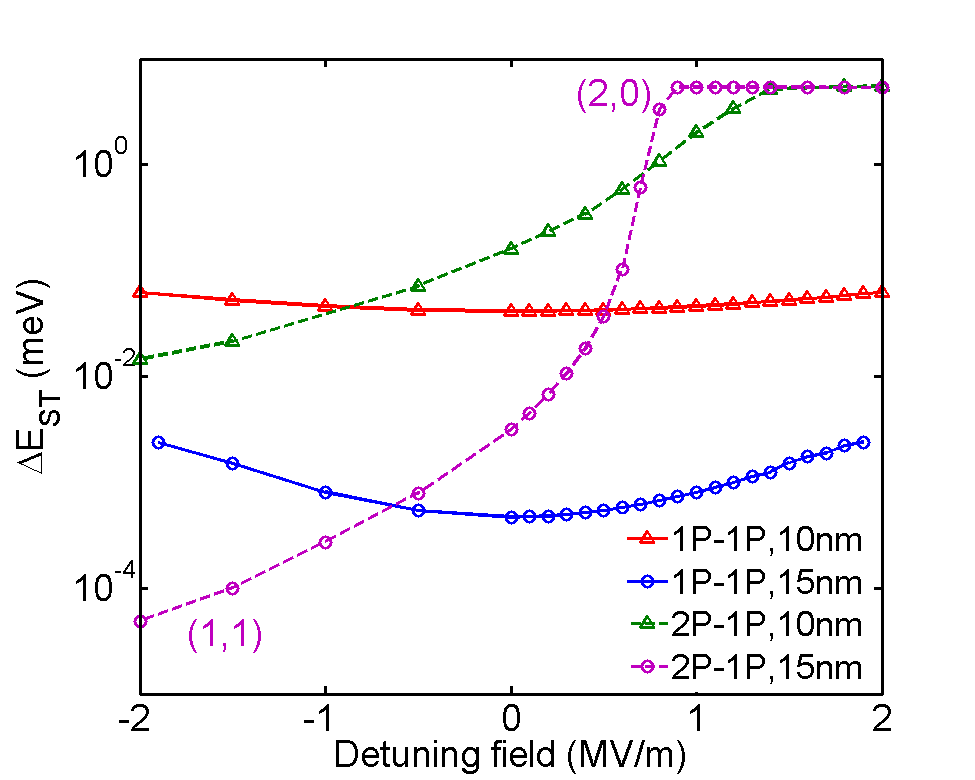}
\caption{ Electric field induced exchange in donors. Exchange energy as a function of a detuning field along the separation axis for a 1P-1P and a 2P-1P system with inter-dot separations of 10 nm and 15 nm. The 2P(B) configuration is used.}
\vspace{0cm}
\label{fi1}
\end{figure}

We note that within the 2 MV/m field range, only the 2P-1P cases exhibit significant exchange tunability in Fig. 3. The 1P-1P case only shows a change from 0.5 $\mu$eV to 2.4 $\mu$eV (5 times) over 2 MV/m for a 15 nm separation distance. Since the 2P cluster has two core nuclear charges with one bound electron, it is easier to shuttle an extra electron to this system aided by the net attractive potential of the core. In the 1P-1P case, the electron-electron repulsive energy is stronger due to the charge neutrality of each qubit, and a larger E-field is needed to reach the (2,0) regime. The calculations show that for E$<2$ MV/m, the (2,0) regime is never reached in the 1P-1P case if the separation is less than 15 nm. In addition, since such a system is symmetric, either a positive or a negative field will increase the exchange energy. It can be seen that qubit pairs with larger separations exhibit larger tunability. This is because the same E-field causes a larger potential drop between the qubits if their separation is larger, and hence provides a larger detuning energy. We also note that the exchange curves show that the transition to the (2,0) regime is smoother if the separation distance is less. This is due to the stronger molecular hybridization of closely spaced qubit pairs. 

After comparing the 1P-1P and 2P-1P cases in Fig. 3, one can see that a 2P-1P system with a 15 nm separation provides a promising two-qubit unit of a silicon quantum computer, yielding 5 orders of magnitude exchange tunability. In our simulations with uniform electric fields, we are unable to go to the high field regime for the 1P-1P case, as high fields induce a triangular quantum well at the lateral domain boundaries causing electron localization in the surface states \cite{Rahman_orbital_stark_effect}. However, larger exchange energies can possibly be realized even in the 1P-1P case with large spatially varying E-fields from detuning gates as shown in Fig. 1(b) and (c). In Appendix, we show that the exchange energy of a 1P-1P system can be tuned by a factor of 50 using surface detuning gates. We also compare the exchange tunability of corresponding cases under the Kane architecture and show the detuning scheme is superior. These simulations include electrostatic simulations with the Sentaurus technology computer-aided design tool (TCAD) \cite{Sentaurus} coupled to the TB-FCI method. 

\subsection{C. Two-electron spectrum}
We now study the promising 2P-1P system for a 15nm separation in more detail including the two-electron energy spectrum and the singlet and triplet wavefunctions at various detuning fields. Fig. 4(a) demonstrates the two-electron total energies of several lowest singlet (black curves) and triplet (blue curves) states as a function of the detuning field. The lowest solid lines are the (1,1) and (2,0) singlet states. They mix at low detuning fields and anti-cross at 0.7 MV/m, which marks the symmetry point of the (1,1) to (2,0) charge transition. The lowest (1,1) triplet state represented by the blue dotted line has little dependence of the detuning field until 0.9 MV/m, indicating that it is decoupled from other triplet states. A sharp transition occurs at 0.9 MV/m, beyond which the energies of singlet and triplet states have a linear dependence on the electric field, indicating the occupation regime of a (2,0) charge configuration where the lowest states of the two qubits are no longer tunnel-coupled. 

\begin{figure}[htbp]
\center\epsfxsize=3.4in\epsfbox{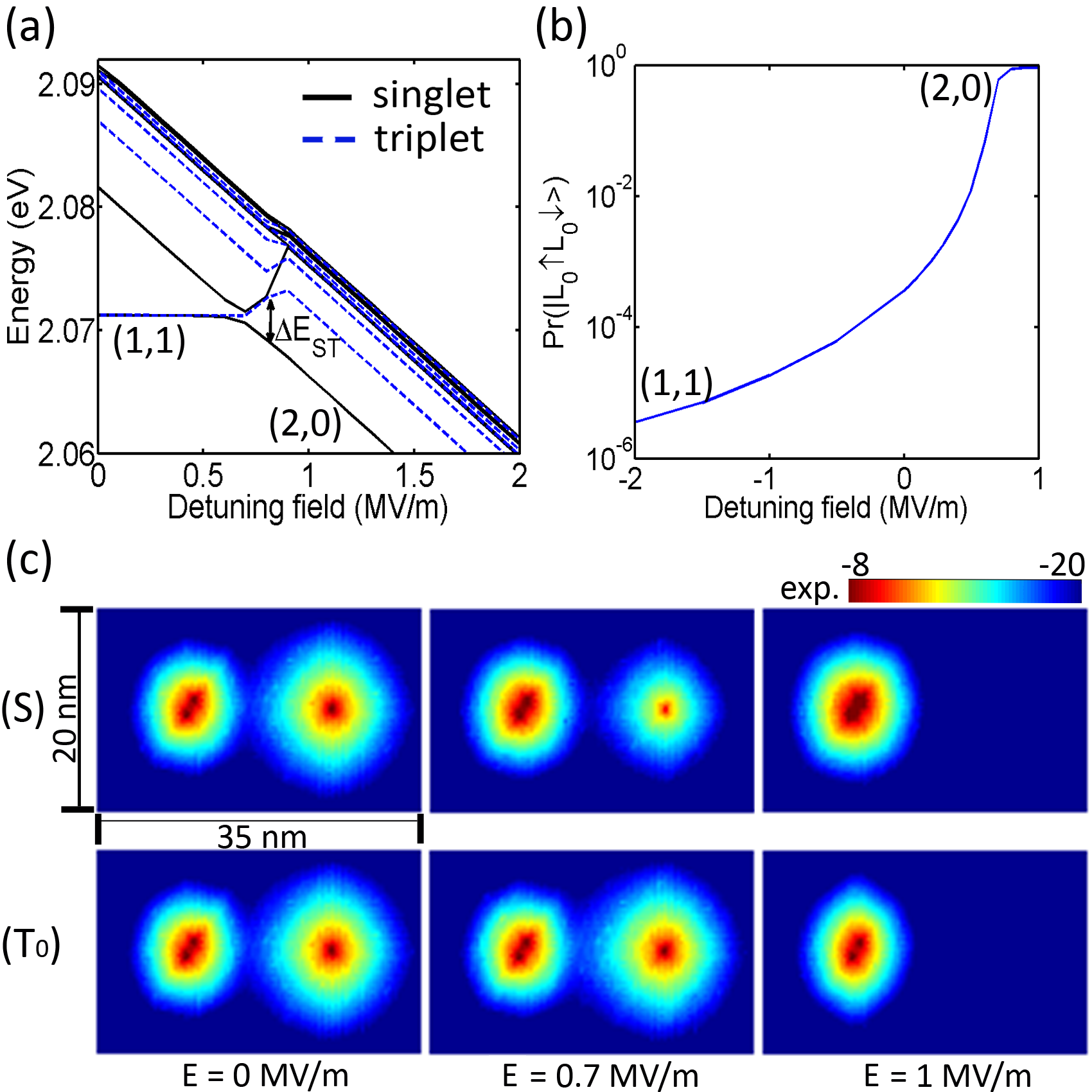}
\caption{ Electric field dependence of singlet and triplet states. (a) The total energy of the lowest few singlet and triplet states of a 2P-1P system with a 15 nm inter-qubit separation as a function of detuning field. (b) Probability of the (2,0) state with both electrons on 2P as a function of detuning. (c) Two electron density of the lowest singlet (S) and triplet (T$_0$) states at three detuning field values, showing a transition from (1,1) to (2,0).}
\vspace{0cm}
\label{fi2}
\end{figure}

Fig. 4 (b) shows the weight of the lowest (2,0) Slater Determinant $\vert$$L_0$$\uparrow$$L_0$$\downarrow$$\rangle$ in the ground state singlet of the 2P-1P system as a function of the detuning electric field. As can be seen, the weight changes from a small probability 3$\times$10$^{-6}$ to $\sim$1, indicating a charge transition from (1,1) to (2,0). This (2,0) probability can be measured by a charge sensor, as in experiments with DQD based singlet-triplet qubits \cite{Petta,Shulman}. Thus the same control scheme of Ref \cite{Petta} can be utilized to realize singlet-triplet based donor qubits with electrical manipulation of J. In such a scheme, the system is prepared in the (2,0) state, pulsed into the (1,1) state, allowed to evolve through the J-coupling, and pulsed back and measured in the (2,0) state - all steps being controlled by the detuning field. 

In Fig. 4(c), we show the two-electron density of the lowest singlet and triplet states computed from the FCI wavefunctions at three different E-fields. At $E=0$, the singlet and triplet states look similar as they are both in the (1,1) charge configuration. The stronger confinement in the 2P cluster on the left is responsible for a smaller wavefunction extent. At $E=0.7$ MV/m, the (1,1) singlet mixes with the (2,0) singlet and the electron density gradually shifts to the 2P cluster. However, due to spin blockade, the triplet still remains in the (1,1) configuration, with almost negligible change in the wavefunction. 
At high enough detuning fields of 1 MV/m, both the singlet and the triplet are in the (2,0) regime, as verified by the electron densities being localized in the 2P cluster.    

In summary, we have shown that a detuning gate control of the J-coupling in donors can relax the stringent requirements of gate widths, gate densities, and precision donor placement needed to realize a two-qubit gate in silicon. In addition, the use of an asymmetric 2P-1P qubit pair can yield a giant tunability of the exchange (5 orders of magnitude) over a modest field range (3 MV/m) with an even lower 'Off' state exchange than the corresponding 1P-1P qubits. Combined with long $T_1$ times of donor clusters \cite{Hsueh}, improved addressability of 2P-1P qubits \cite{Buch}, and operation schemes of Ref \cite{Rachpon}, the proposed design helps in the experimental realization of the much sought after two-qubit gate with donors in silicon. The calculations of detuning controlled exchange energy also helps in highly tunable realizing singlet-triplet electronic qubits along the lines of DQDs \cite{Petta}. 

     
\section*{ACKNOWLEDGEMENTS}
This research was conducted by the Australian Research Council Centre of Excellence for Quantum Computation and Communication Technology (project No. CE110001027), the US National Security Agency and the US Army Research Office under contract No. W911NF-08-1-0527. Computational resources on nanoHUB.org, funded by the NSF grant EEC-0228390, were used. M.Y.S. acknowledges a Laureate Fellowship. This research is part of the Blue Waters sustained-petascale computing project, which is supported by the National Science Foundation (award number ACI 1238993) and the state of Illinois. Blue Waters is a joint effort of the University of Illinois at Urbana-Champaign and its National Center for Supercomputing Applications. This work is also part of the “Accelerating Nano-scale Transistor Innovation with NEMO5 on Blue Waters” PRAC allocation support by the National Science Foundation (award number OCI-0832623).

\section*{APPENDIX}
We demonstrate the exchange tunability of the 1P-1P and 2P-1P systems in realistic gated device structures as shown in Fig. 1. Comparison is made between the Kane J-gate (Fig. 1(a)) and the detuning gate structures (Fig. 1(b) and (c)). For the former case, $A$ represents the hyperfine gates and $J$ the exchange gates, whose width are 6 nm. The dark red region under these gates is 3 nm thick SiO$_2$ and the light pink region is undoped silicon. The qubit-donors are placed right beneath the middle of the two A-gates respectively, separated by 15 nm and $\sim$10 nm below the Si/SiO$_2$ interface. 0 V is applied on the A-gates and the middle J-gate bias is varied from -1 V to 1 V to tune the exchange energy. The electrostatic potential of the device is solved by a technology computer aided design tool Sentaurus \cite{Sentaurus} and then input to the atomistic TB-FCI calculations. 

\begin{figure}[htbp]
\renewcommand\thefigure{S}  
\centering\epsfxsize=3.4in\epsfbox{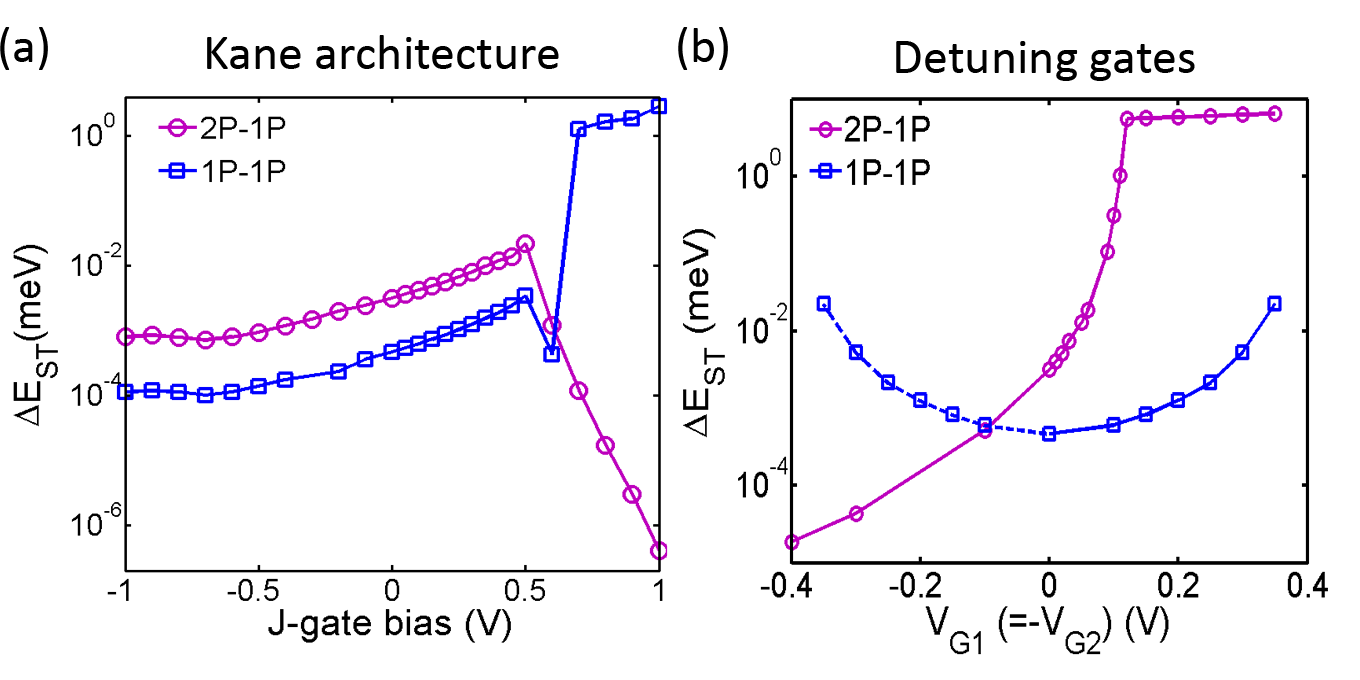}
\caption{ Comparison of tuning exchange energy with Kane's J-gate and detuning gates. (a) Exchange energy as a function of J-gate biases for a 1P-1P and a 2P-1P systems with an inter-qubit separation of 15 nm along the [100] direction. The device structure is shown as Fig. 1(a). (b) Exchange energy as a function of top-gate biases for a 1P-1P and a 2P-1P systems with an inter-qubit separation of 15 nm along the [100] direction. The device structure is shown as Fig. 1(b).}
\vspace{0cm}
\label{fis}
\end{figure}

Fig. S(a) shows the exchange energy of the 1P-1P and 2P-1P systems as a function of the J-gate bias. By applying negative biases, the exchange energy first decreases between 0 V and -0.6 V because the two electron are pushed further apart such that their wavefunction overlap diminishes. Then the exchange energy reaches a plateau between -0.6 V and -1 V because the electronic wavefunctions are already pushed as far apart as possible by the J-gate bias, resulting in a constant overlap. By applying positive biases, the exchange energy increases between 0 V and 0.5 V as the two- electron wavefunctions are pulled towards the middle region underneath the J-gate, and their overlap increases as a result. By further increasing the J-gate bias, for the 1P-1P system, the exchange energy first decreases (0.6 V) as one of the donor electron is ionized to the interface while the other is still at the donor site (depending on whether the 6 nm J-gate has a bit of asymmetry with respect to the donors). Then the exchange energy drastically increases (0.7 V to 1 V) because both electrons are ionized to the interface and their wavefunction overlap increases sharply. However, in this regime, the interface electrons are not distinguishable any more and cannot be used as qubit electrons. 

For the 2P-1P system, the exchange energy decreases from 0.6 V to 1 V because at 0.6 V, the 1P electron is ionized to the interface but the 2P electron is still bound by the 2P donors. The interface electron is prone to interface charges and gate noise, which can significantly undermine the coherence time and the qubit performance. So this regime for the 2P-1P system is not favorable, either. Thus, the exchange can be tuned only by a factor of ~30 (-1 V to 0.5 V) under Kane's architecture for both the 1P-1P and the 2P-1P systems placed 10 nm deep into the lattice. Since the ionization depends on whether the surface states near the J-gates are lower in energy than the donor states, the ionization bias can vary with donor depth. In donors closer to the surface, the electron can remain donor bound at larger fields or gate biases \cite{Rahman_orbital_stark_effect}.       

In comparison, we also investigated the exchange tunability with top and in-plane detuning gates. In Fig. 1(b), G$_1$ and G$_2$ are two top gates with a width of 6 nm. The qubit donors are placed beneath the center of G$_1$ and G$_2$, respectively, separated by 15 nm and $\sim$10 nm below the Si/SiO$_2$ interface. Opposite biases with equal magnitudes are applied on G$_1$ and G$_2$ to detune the system. Fig. S(b) displays the exchange energy as a function of the detuning gate biases for both the 2P-1P and the 1P-1P systems. For 2P-1P, the Fig. S(b) demonstrates a similar exchange tunability, 5 orders of magnitude, to the uniform field detuning scheme as shown in Fig. 3 of the main text. The in-plane detuning gate structure (Fig. 1(c)) provides similar tunability as well (not shown). For the 1P-1P system, larger spatially varying electric fields applied from surface gates increases the J-tunability by a factor of ~50 (from 0.5 $\mu$eV to 22.1 $\mu$eV) as shown in Fig. S(b). Compared with a factor of 5 tunability with the uniform field detuning scheme, detuning with top-gates shows a more promising tunability of the inter-donor exchange energy. The exchange tunability in the 1P-1P system will therefore be limited by the magnitude of the electric field realizable in practice. Better performance could possibly be achieved with more rigorous device geometry and control optimization. 

In summary, the J-gate in Kane's architecture provides limited exchange tunability for both 1P-1P and 2P-1P systems. Detuning gate structures are better designs that provide larger exchange tunability. This work therefore provides guidance to two-qubit device designs for future experiments.

Electronic address: wang1613@purdue.edu, rrahman@purdue.edu


\begin{thebibliography}{100}
\bibitem{charge_config_note} The charge configuration of the system is denoted as (M,N), where M indicates the number of electrons on the left dot, and N the right dot.
\end{thebibliography}
\end{document}